\documentclass[aps,twocolumn,superscriptaddress,prl]{revtex4-2}
\usepackage[utf8]{inputenc}
\usepackage{amsmath, amssymb, graphicx, color, subfigure}
\usepackage[bookmarks=false]{hyperref}
\usepackage[usenames,dvipsnames]{xcolor}
\usepackage[super]{nth}
\usepackage[normalem]{ulem}
\usepackage[dvipsnames]{xcolor}

\def\be{\begin{equation}}
\def\ee{\end{equation}}
\def\ba{\begin{eqnarray}}
\def\ea{\end{eqnarray}}
\newcommand{\blue}[1]{\textcolor{blue}{#1}}

\newcommand{\Brown}[1]{\textcolor{Brown}{#1}}

\begin{document}

\title{Anomalous metals: From ``failed superconductor'' to ``failed insulator''}

\author{Xinyang Zhang}
\affiliation{Geballe Laboratory for Advanced Materials, Stanford University, Stanford, CA 94305, USA}
\affiliation{Department of Applied Physics, Stanford University, Stanford, CA 94305, USA}

\author{Alexander Palevski}
\affiliation{School of Physics and Astronomy, Raymond and Beverly Sackler,
Faculty of Exact Sciences, Tel Aviv University, Tel Aviv 6997801, Israel}

\author{Aharon Kapitulnik}
\affiliation{Geballe Laboratory for Advanced Materials, Stanford University, Stanford, CA 94305, USA}
\affiliation{Department of Applied Physics, Stanford University, Stanford, CA 94305, USA}
\affiliation{Department of Physics, Stanford University, Stanford, CA 94305, USA}

\date{\today}

\begin{abstract}
Resistivity saturation is found on both superconducting and insulating sides of an ``avoided'' magnetic-field-tuned superconductor-to-insulator transition (H-SIT) in a two-dimensional In/InO$_x$ composite, where the anomalous metallic behaviors cut off conductivity or resistivity divergence in the zero-temperature limit. The granular morphology of the material implies a system of Josephson junctions (JJ) with a broad distribution of Josephson coupling $E_J$ and charging energy $E_C$, with a H-SIT determined by the competition between $E_J$ and $E_C$. By virtue of self-duality across the true H-SIT, we invoke macroscopic quantum tunneling effects to explain the temperature-independent resistance where the ``failed superconductor'' side is a consequence of phase fluctuations and the ``failed insulator'' side results from charge fluctuations. While true self-duality is lost in the avoided transition, its vestiges are argued to persist, owing to the incipient duality of the percolative nature of the dissipative path in the underlying random JJ system.

\end{abstract}

\maketitle

\section{Introduction}
An increasing number of recent experiments on disordered superconducting films, initially searching for a superconductor-to-insulator transition (SIT),  end up instead with an ``avoided'' transition, where scaling about a putative zero-temperature quantum critical point fails in the limit $T\to 0$, giving way to a phase characterized by a resistance that levels off to a constant value independent of temperature  \cite{Ephron1996, Chen1996, MasonKapitulnik1, Breznay2017,Bottcher2018, Chen2018,Yang2019, Zhang2021}.  This ``anomalous metallic state'' (AMS) seems to exhibit electronic properties that cannot be understood on the basis of the standard paradigms for transport in disordered two-dimensional (2d) metals (for a recent review see \cite{KKS2019}).   While extensive efforts have been devoted to understanding AMS on the superconducting side, termed a ``failed superconductor,'' \cite{KKS2019} and its transition to a ``true'' superconducting state \cite{MasonKapitulnik2}, the complementary leveling of resistance on the insulating side of the same avoided SIT is often overlooked. Here, an initial resistivity divergence trend for $T\to 0$ is replaced with 
distinct saturation, exhibiting resistivity values that can be orders of magnitude higher than quantum of resistance for Cooper pairs $R_Q =h/e^2\sim 6.45\ {\rm k}\Omega$. This ubiquitous behavior of a ``failed insulator'' can be found in studies of ultrathin granular metal films \cite{Jaeger1986}, amorphous metal-insulator films \cite{MasonKapitulnik1,Couedo2016}, and Josephson junction (JJ) arrays \cite{Delsing1994}. 

In this letter we propose a unified phenomenology for the emergence of anomalous metallic states from an avoided SIT in disordered superconducting films. Focusing on the magnetic-field tuned SIT (H-SIT) in strongly granular system of In-on-InO$_x$ composites, we show that vestiges of the self-duality observed in a true H-SIT \cite{Breznay2016,Hen2021} are still pronounced in the AMS on \textit{both} sides of the avoided-H-SIT \cite{Zhang2021}, including the resistance saturation. Bearing in mind that from its nature, the critical point of any SIT yields an anomalous metal, we propose that the broadening of the metallic phase on both sides of the H-SIT has the same origin modulo the dual relation between phase of the order parameter (or vortices) and charge (both cooper-pairs and single electrons). Specifically, we argue that the low-temperature resistance  can be interpreted as a sum of a temperature-dependent activation term and a temperature-independent term associated with quantum fluctuations leading to macroscopic quantum tunneling (MQT) of the phase and/or the charge. On the superconducting side, a superconductor-to-quantum-metal transition (SQMT) appears with the destruction of global phase coherence \cite{Hruska,OretoKivSp,KKS2019}, while on the insulating side quantum charge fluctuations prevent the establishment of a Coulomb-blockade-driven insulating state \cite{Averin1986,Fazio1991,Delsing1994} driving a quantum-metal-to-insulator transition (QMIT). 

\begin{figure}[h]
	\centering
	\includegraphics[width=1.0\columnwidth]{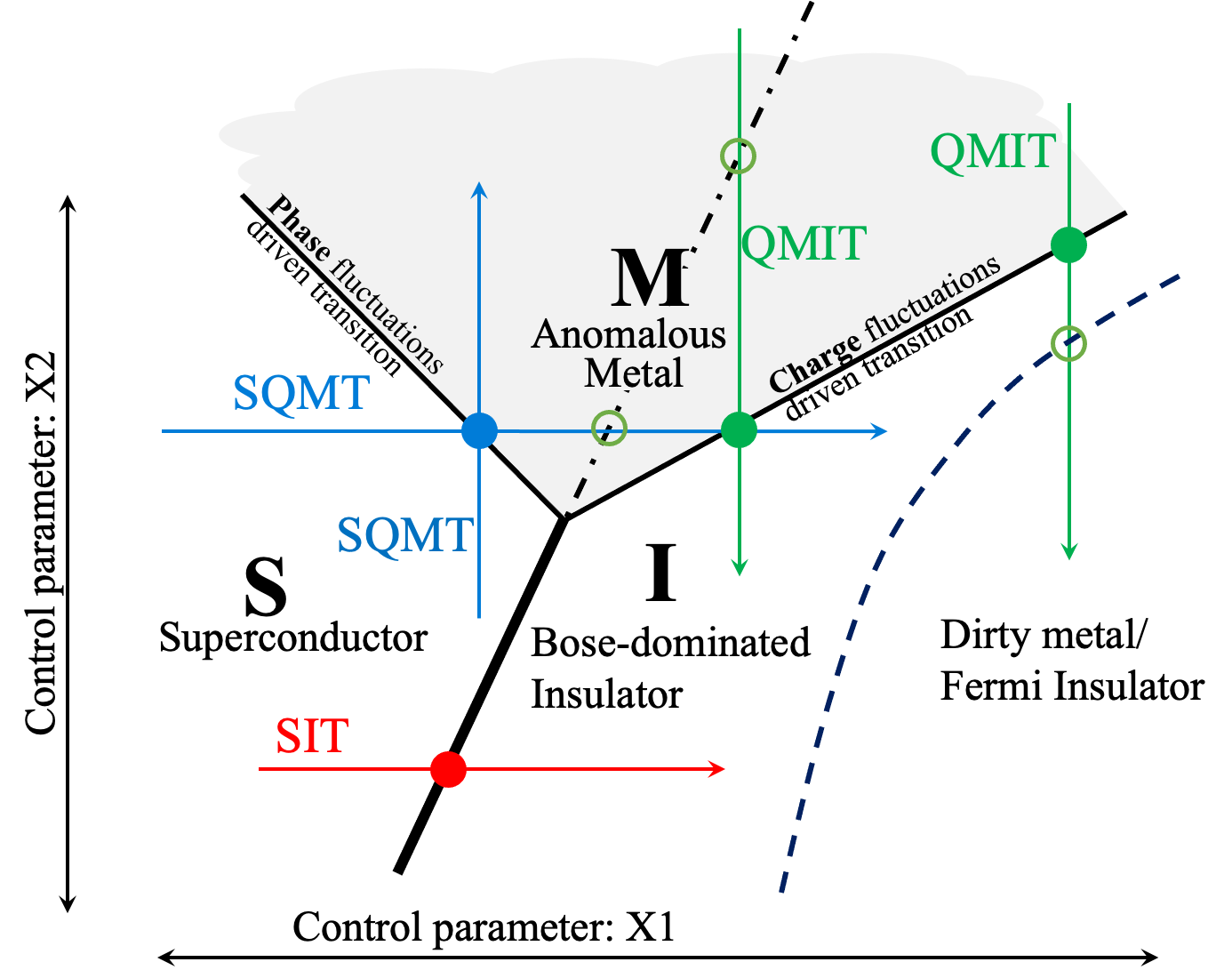}
	\caption{Zero-temperature phase diagram of a 2d granular superconductor with two controlling parameters. In the present study X1=magnetic field and X2=inter-grain coupling. Shaded area represents the region in parameter space where anomalous metal appears. Thick solid line represents a true H-SIT (Ref.~\cite{Hen2021}) while the other solid lines represent the transition from a superconductor to a quantum anomalous metal (SQMT) and from quantum anomalous metal to an insulator (QMIT). Dashed line represents crossover from a Bose-dominated system, where superconducting In islands provide the source of pairing, to a normal-electrons insulator.  Dash-dotted line represents the crossover from phase to charge dominated quantum anomalous metal state. Each full circle represents a quantum critical point for a particular realization of the control parameters, while an open circle is a crossover point.}
	\label{fig1}
\end{figure}

The robustness of the anomalous metallic phase in the In/InO$_x$ system \cite{Zhang2021} further enforces these expanded observations, while strongly contending interpretations solely based on non-equilibrium effects and response to external effects of the electronic system \cite{Tamir2019} (see also ``Methods'' section). At the same time, the  temperature-independent term associated with MQT depends delicately on the details of the distribution of grains and junctions, which may be controlled by external effects leading to different saturation value of the resistance, emphasizing its non-universal value in the anomalous metal regime. This in turn leads us to propose the phase-diagram of Fig.~\ref{fig1} and suggest a unified understanding of anomalous metallic states in disordered superconducting films. Excluding the regime where pairs are no longer present, this phase diagram has strong resemblance to that of two-dimensional electron gas about half-filled Landau levels \cite{Mulligan2016}. 

Whether samples are inherently inhomogeneous such as arrays of Josephson-junctions (JJ), or morphologically uniform such as amorphous films,  superconducting pairing interaction tends to ``amplify'' mesoscopic fluctuations of the disorder and its associated Coulomb interaction, resulting in an effective granular morphology \cite{Shimshoni1998,Ghosal1998,Skvortsov2005,Dubi2007}. Therefore, it is reasonable to assume that the superconducting transition in 2d disordered metallic films is dominated by phase fluctuations \cite{LarkinFeigelman,Hruska,OretoKivSp} and the material can be modeled as superconducting grains embedded in a tunable intergrain coupling matrix. While local pair amplitude fluctuations depend on the grain size \cite{Muhlschlegel1972}, global superconductivity is achieved via the establishment of a percolating path of phase coherence \cite{Entin1981,Ioffe1981,ImryStrongin,Merchant2001} through Josephson coupling of pairs of adjacent grains. For a given pair $\{ij\}$, $E_J^{ij}$ is a function of the local gaps $\Delta_i$ and $\Delta_j$ and the intergrain normal-state resistance $R_n^{ij}$. However,  the granular nature of 2d disordered films also requires that we consider the capacitance (both  intergrain and self) involved and the associated charging energy $E_C^{ij}$ which is a function of the local density at the grains $n_i$ and $n_j$, and the dielectric constant of the intergrain material $\chi_n^{ij}$. An insulating phase is the result of Coulomb blockade which prevent Cooper-pairs tunneling between adjacent grains.  For a granular film with typical Josephson energy $E_J$ and charging energy $E_C$, the ratio $\gamma \equiv E_J/E_C$ determines the occurrence of SIT \cite{Abeles1977,Efetov1980,Fazio1991}. Besides the initial film morphology, other external parameters such as an applied magnetic field or carrier density modulation through an applied gate can be used to control $\gamma$.

\section{Methods}
\begin{figure}[h]
	\centering
	\includegraphics[width=1.0\columnwidth]{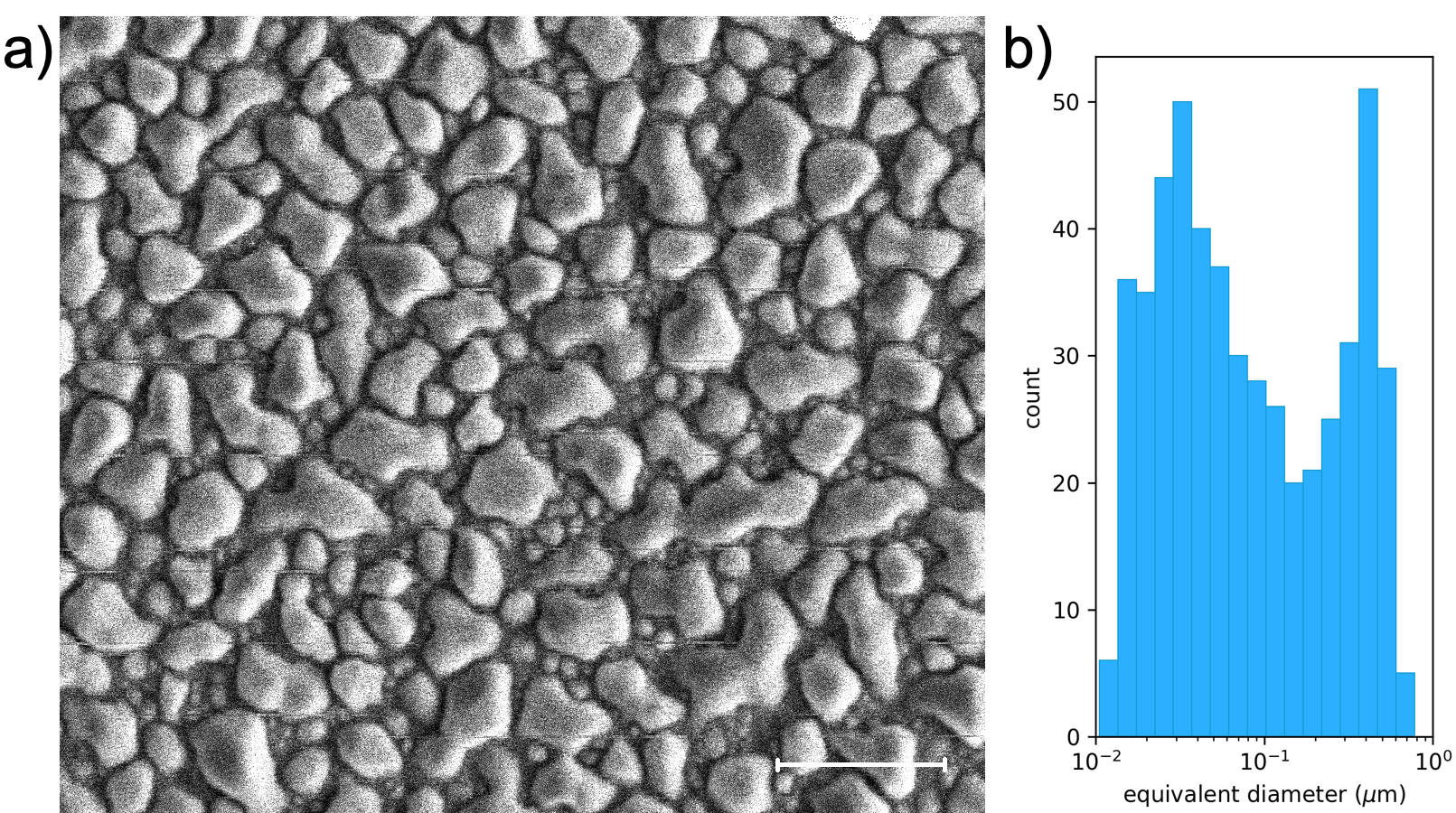}
	\caption{Morphology of the In/InOx system: a) Scanning electron microscopy (SEM) micrograph of the In/InOx sample. Bright grains are metallic indium islands, while dark background is a uniform underlying layer of weakly-insulating amorphous InOx. Note the broad distribution of grain sizes including the interstitial ones. Scale bar at lower right corner indicates 1$\mu$m. b) Distribution of grain size. Note the two broad peaks at $\sim 0.03 \mu$m and $\sim 0.5 \mu$m. }
	\label{fig2}
\end{figure}

InO$_x$/In granular composites were grown by electron-beam evaporation of In$_2$O$_3$ in oxygen partial pressure, followed by that of In under high vacuum. Consecutive \textit{in situ} depositions of the two components result in clean interface that is crucial for optimal interface coupling. The uniform underlying amorphous InO$_x$ is specifically prepared to be weakly insulating to mediate couplings among the In grains and nearby proximitized regions. Details of the sample preparation process were provided in Ref.~\cite{Zhang2021}. Here we emphasize the granular morphology of the system as shown in Fig.~\ref{fig2}. This will be a key to our understanding of the data.

Resistivity measurements were carried out using standard four-point lock-in measurement on a patterned hall-bar sample. Current excitation was provided by reference ac voltage source in series with a 1\ G$\Omega$ resistor that is much more resistive than any measured sample resistance in this study. The current intensity ranged from 0.1\ nA to 1\ nA while the frequency varied between 0.7--7\ Hz depending on sample resistance. Linear response to excitation current was confirmed at the highest resistance. Temperature ramps were sufficiently slow compared to the prolonged time constant. 

In our dilution refrigerator, all measurement wires were filtered at both room temperature and mixing chamber temperature ($T\gtrsim10$ mK) using commercial in-line pi-filters and cryogenic filters, achieving over 100 dB attenuation throughout 100\ MHz--5 GHz range. Sample phonon temperature was measured by a calibrated on-chip RuO$_2$ thermometer subject to the same mounting and wiring as the sample. Data is not considered in our analysis when sample temperature differs more than 10\% from mixing chamber temperature. The refrigerator was operating at the highest cooling power with an extra circulation pump running, despite at the cost of a higher base (mixing chamber) temperature.

\section{Results}
\begin{figure}[h]
	\centering
	\subfigure{\label{fig3a}\includegraphics[width=0.365\columnwidth]{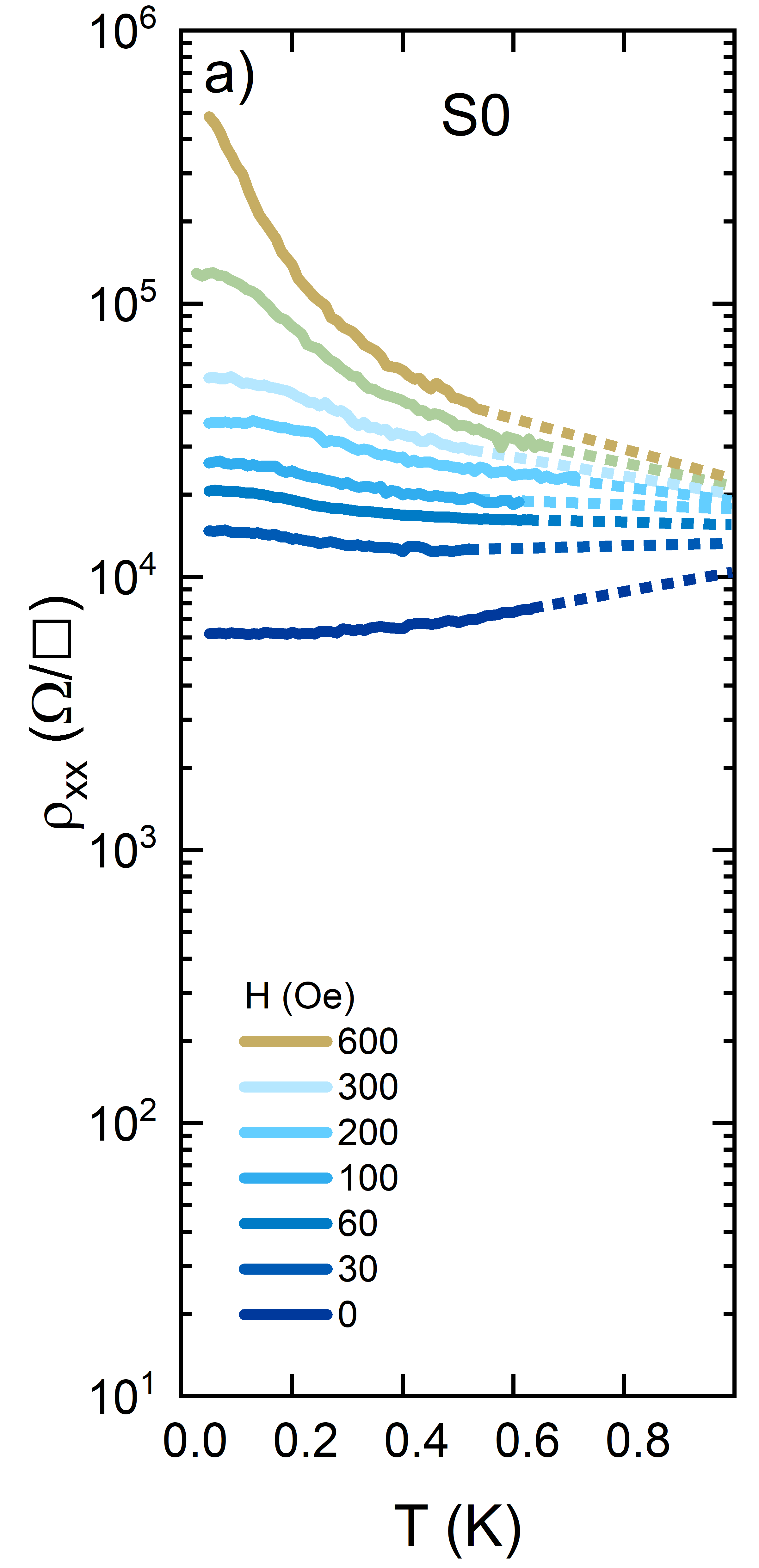}}
	\subfigure{\label{fig3b}\includegraphics[width=0.292\columnwidth]{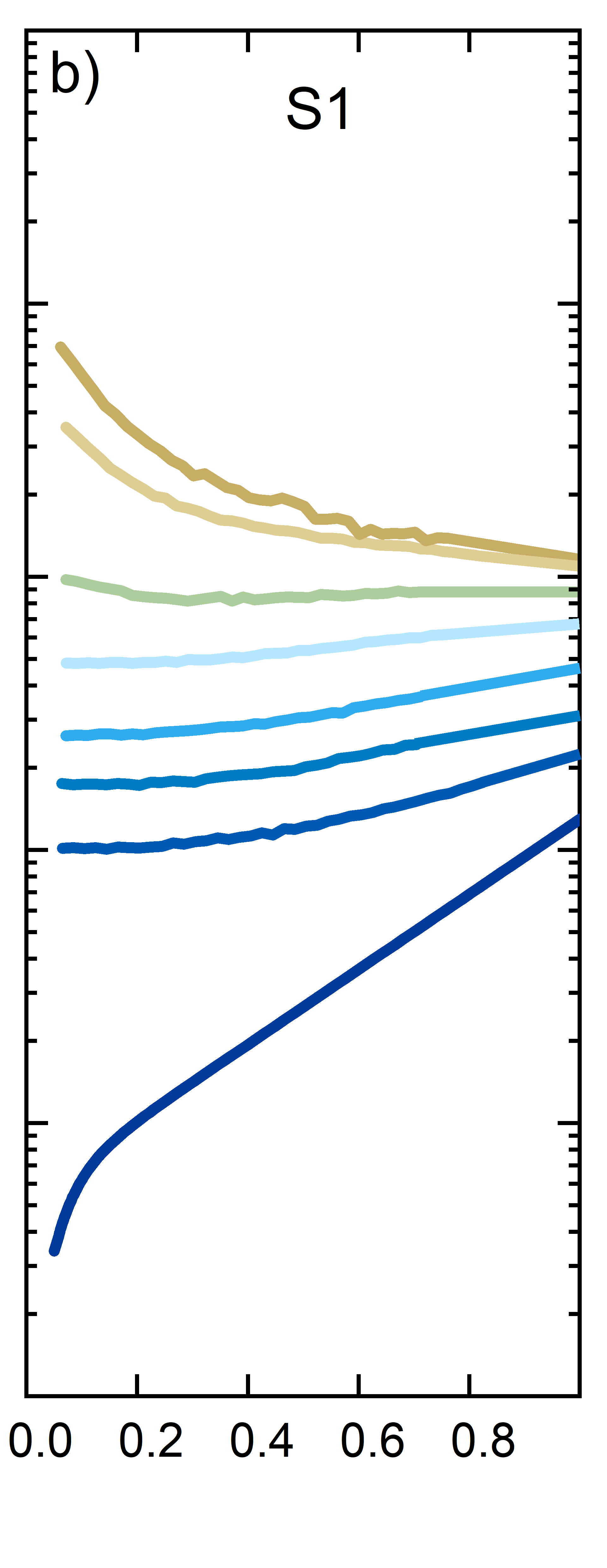}}
	\subfigure{\label{fig3c}\includegraphics[width=0.319\columnwidth]{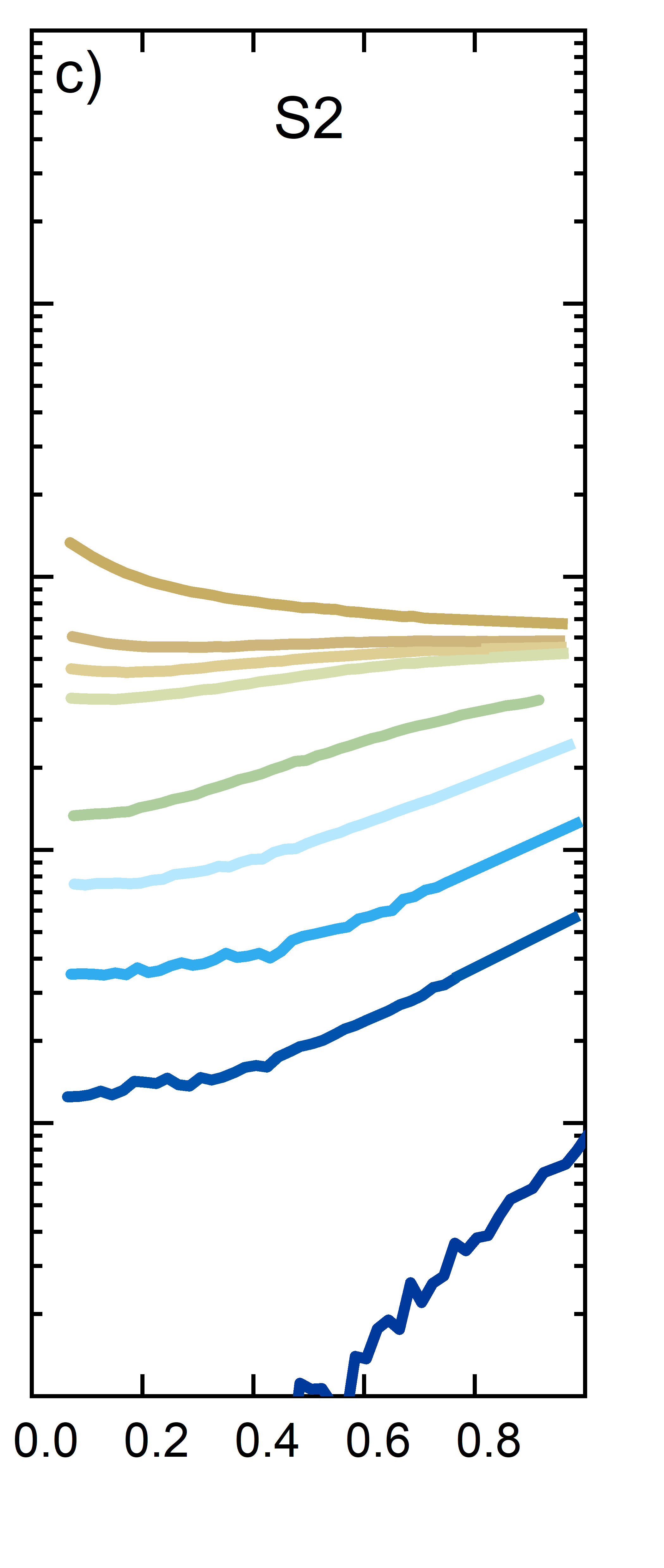}}
	\caption{Resistivity vs. temperature  in varying magnetic field for three successive annealing stages. Dotted lines in S0 connects cooling data $<0.7$\ K with a field sweep at 1.1\ K.}
	\label{fig3}
\end{figure}

In Fig.~\ref{fig3}, we plot temperature dependence of resistivity in different perpendicular magnetic fields across three annealing stages S0, S1, and S2 of the same initial sample. Low-temperature annealing ($<50\ ^\circ$C) is well-known \cite{KowalOvadyahu} to irreversibly reduce sample resistance yet maintain an amorphous nature of the underlying InO$_x$, thus altering the dielectric response in the insulating matrix that couples superconducting grains. We restrict the range of magnetic field in this analysis to under 600\ Oe to ensure robust superconductivity within the single grain \cite{Hen2021}.

Annealing stage S0 shows the most salient signatures of resistivity saturation on the insulating side of an ``avoided'' SIT. In zero field as temperature is lowered, resistivity decreases and saturates at around 6.2\ k$\Omega/\square$ as a failed superconductor. A magnetic field as low as 30\ Oe  flips temperature coefficient of resistance (TCR) to negative as in an insulator. However, the resistivity divergence gives way to saturation at around 0.2\ K, eventually settling at a finite value as $T\rightarrow0$. In higher fields the features are qualitatively similar, until at 600\ Oe  where resistivity strongly diverges and no saturation is found above our base temperature. In a broad range of magnetic field, we identify an insulator where resistivity fails to diverge as $T\rightarrow0$, dubbed as a ``failed insulator.''

\begin{figure}[h]
	\centering
	\subfigure{\label{fig4a}\includegraphics[width=\columnwidth]{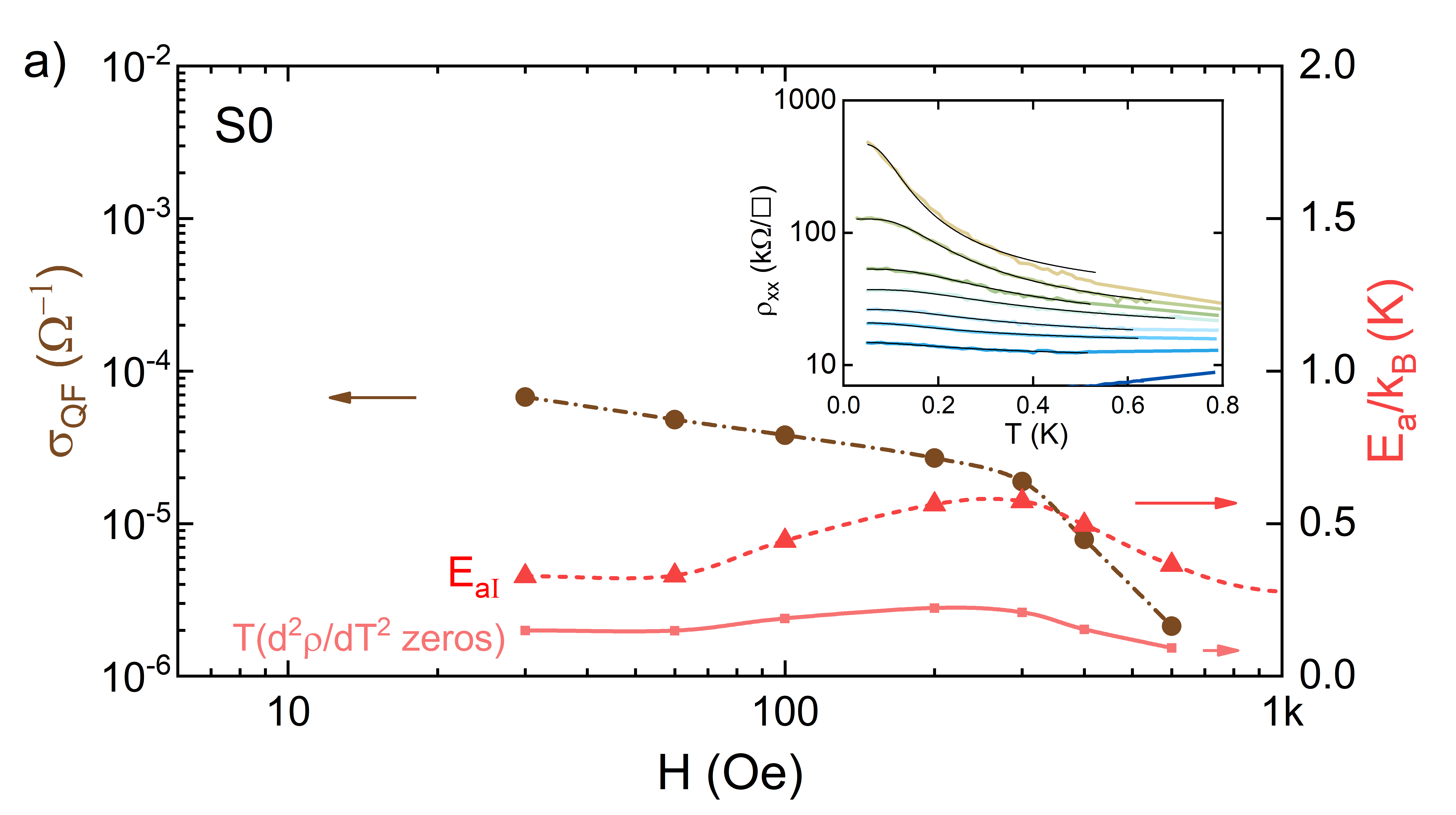}}
\subfigure{\label{fig4b}\includegraphics[width=\columnwidth]{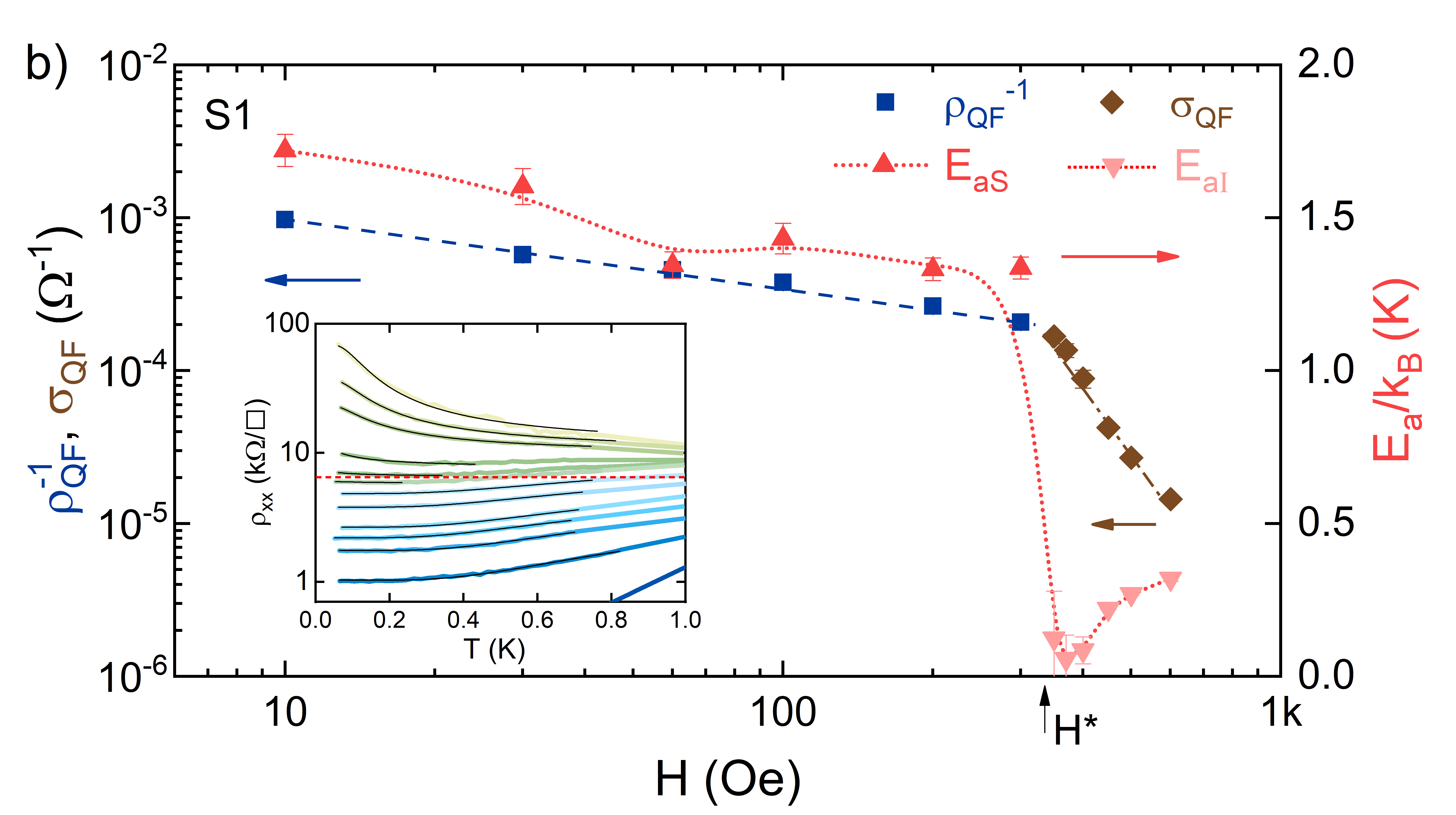}}\\
	\caption{\textbf{a)} Temperature independent term $\sigma_{QF}$ (left axis - \Brown{$\bullet$}) and activation temperature $E_a/k_B$ (right axis) extracted from the fits. Also shown are the zeros (light red) in the curvatures of $\rho(T)$. Inset: Temperature dependence of resistivity in S0 fitted by empirical law Eq.~\ref{eq1}.  \textbf{b)} Temperature independent term $\rho_{QF}^{-1}$ (\blue {$\blacksquare$}) or $\sigma_{QF}$ (\Brown{ $\blacklozenge$}) (left axis) and activation temperature $E_a/k_B$ (right axis) extracted from the fits. Inset: Temperature dependence of resistivity in S1 fitted by empirical laws Eq.~\ref{eq2}. for $H\leq H^*$ and Eq.~\ref{eq1}. for $H\geq H^*$. Also shown is Cooper pair quantum of resistance $R_Q=h/4e^2\sim 6.5\ \text{k}\Omega/\square$ as the red separatrix. Dotted and dashed lines are guidance to the eye.}
		\label{fig4}
\end{figure}

In contrast, annealing stage S1 or S2 develops a superconducting ground state in zero field, where the $T_c$ is just below our base temperature for S0 while $T_c\approx0.4$\ K for S1. Nevertheless, in low field $\lesssim400$\ Oe, the ground state is a failed superconductor and has been discussed extensively \cite{Zhang2021}. For S1 at 400\ Oe, the system is right on the insulating side of the transition, yet the resistivity is saturating as $T\rightarrow0$ despite an upturn at $\sim0.2$ K. In higher field $\gtrsim400$\ Oe for both S1 and S2, resistivity saturation is substantially reduced and the temperature dependence can be fitted by a logarithmic divergence with a large pre-factor \cite{Zhang2021}. Such logarithmic divergence is in stark contrast with the saturated resistivity in S0. Resistivity of insulators typically have exponential (hopping) or weak, granularity-induced divergence persisting to zero-temperature, meaning their curvature $d^2\rho/dT^2$ remains positive as $T\rightarrow0$. However, in S0 the resistivity divergence is cutoff by a saturation, marked by a sign change in the curvature in the temperature dependence at around 0.2\ K (see Fig.~\ref{fig4a}).

To quantitatively analyze the evolution of resistivity behavior as a function of magnetic field on the \textit{insulating} side, we adopt an empirical functional form Eq.~\ref{eq1}, which satisfactorily fits S0 data as shown in Fig.~\ref{fig4a}. This empirical law was previously employed \cite{Delsing1994} to describe an analogous conductivity saturation in insulating JJ arrays. The saturation was modeled as a consequence of a temperature-independent quantum fluctuation term $\sigma_{QF}$, in addition to thermally-activated tunneling conductivity on the insulating side with activation energy $E_{aI}$ and a proportionality constant $\sigma_0$. In writing this equation we assume hall conductivity is zero:
\begin{equation}
	\rho=[\sigma_{QF}+\sigma_0\cdot\exp(-E_{aI}/k_BT)]^{-1}
	\label{eq1}
\end{equation}

Fig.~\ref{fig4a} shows an overlay of the extracted quantum fluctuation contribution $\sigma_{QF}$ (left axis) and activation temperature $E_a/k_B$ (right axis) as a function of magnetic field. Since $\sigma_{QF}$ is intimately related to the saturation, it decreases as field increases until around 600\ Oe, where the divergence is restored and the saturation disappears. Meanwhile, the activation temperature starts off at around 0.3\ K, peaks at just below 0.6\ K near 300\ Oe, and subsequently decreases before leveling off at high fields. Since the overall tendency for an insulating behavior is governed by the charging energy, we may estimate $E_{aI}\approx \alpha E_C$, where $\alpha\approx 1/\langle z\rangle$ and $\langle z \rangle \approx 6$ is the average coordination number of the random array of indium grains of average size $\sim 0.1 \ \mu$m \cite{Zhang2021}. Taking a dielectric constant of $\sim10$ for the InOx, and average distance between grains of $\sim 10$ nm, we obtain $E_{aI}$ in the range that we find experimentally in Fig.~\ref{fig4a}. The corresponding $(\sigma_{QF})^{-1}$ increases in the high field limit to order of $\sim 1$ M$\Omega$, emphasizing the fact that this saturated resistance is much larger than the quantum of resistance $R_Q$.

To quantitatively analyze resistivity saturation on the \textit{superconducting} side, we adopt a similar empirical law by exchanging the role of resistivity $\rho$ and conductivity $\sigma$, justified as a manifestation of charge-phase correspondence (or particle-vortex duality) near either a direct SIT or an ``avoided'' SIT -- a crossover between failed superconductors and failed insulators \cite{Breznay2016, Zhang2021}. 
\begin{equation}
	\rho=\rho_{QF}+\rho_0\cdot\exp(-E_{aS}/k_BT)
	\label{eq2}
\end{equation}
where $\rho_{QF}$ is an analogous temperature-independent contribution to resistivity and $E_{aS}$ is activation energy on the superconducting side. Fits of this form to resistivity of S1 are shown in Fig.~\ref{fig4b}, where Eq.~\ref{eq1} is used for $H\geq350$\ Oe  and Eq.~\ref{eq2} for $H\leq300$\ Oe. The left y-axis shows $\rho_{QF}^{-1}$ from low field and $\sigma_{QF}$ from high field in the same plot. Quantum fluctuation contribution $\rho_{QF}$ grows (or equivalently, $\rho_{QF}^{-1}$ shrinks) upon increasing field, while quantum fluctuation contribution $\sigma_{QF}$ shrinks in higher fields. The field dependence for both branches are power laws with $\rho_{QF}^{-1}\propto H^{-0.45}$ and $\sigma_{QF}\propto H^{-4.7}$. Activation temperature on the low field side is around 1.5\ K and gradually decreasing, until a sharp drop at $H^*\sim 300$ Oe  where the sample switches to insulating behavior. Besides a factor of $\sim$4 lower activation temperature, its qualitative trend in field is consistent with those in S0. The dome-shaped presence of $\rho_{QF}$ and $\sigma_{QF}$ suggests that quantum fluctuations play important roles in both regimes of anomalous metallic phase.

\section{Discussion}

The evolution of zero-temperature behaviors of resistivity in S0 and S1 leads us to a holistic understanding of the $T=0$ phase diagram in Fig.~\ref{fig1}. We identify the blue trace (SQMT) as a magnetic-field-tuned quantum superconductor--anomalous-metal transition (solid line), followed by a transition to an insulating ground state (solid line). The anomalous metallic regime can be further divided into failed-superconductor and failed-insulator regimes, separated by an ``avoided'' SIT (dash-dot line) representing a wide crossover regime. We also identify the green trace as an annealing-tuned failed-insulator--failed-superconductor transition (QMIC), similar to the magnetic-field controlled scenario despite a fundamental difference in the nature of these two control parameters. As an example, universal anomalous metallic behaviors under different tuning parameter was recently discussed in Ref.~\cite{Chen2021}. Finally, a direct SIT that is typically found in homogeneous films or granular materials with stronger coupling is shown as the red trace (SIT), while  transition to a disordered 2D metallic phase appears when Cooper-pairs are broken (dashed line).  Within our granular system of In/InOx composite, a true H-SIT \cite{Hen2021} can be tuned to unveil an avoided H-SIT, which was previously shown to exhibit vestiges of self duality around the putative quantum critical point  \cite{Zhang2021}.  The analysis presented in the previous section explores the duality idea in a wider temperature range which includes the resistance saturation associated with the AMS, thus provides further insight into the possible origin of this enigmatic phase.
The fact that the data shows a continuous evolution from an AMS with very low sheet resistance to one with very large sheet resistance that far exceeds even the fermion quantum of resistance, $h/e^2$, suggests that the superconducting grains and the intergrain tunneling play the key role in determining the saturation value, that is, $(\sigma_{QF})^{-1}$  in Eqn.~\ref{eq1} smoothly connects to $\rho_{QF}$ in Eqn.~\ref{eq2}. Focusing on Fig.~\ref{fig4b}, it shows that the separating saturation resistance between the two regimes is $\sim 6.5$ k$\Omega/\Box$ at the field of $H^* \approx 300$ Oe where the curves change character as a result of the avoided H-SIT. The drop in activation energy between $|E_a|_{H\lesssim H^*}$ and $|E_a|_{H\gg H^*}$ is a factor of 4. Since $H^*$ marks an avoided transition where we expect $E_J\sim E_C$, then we conclude that $E_{aI}$ saturates to $\tfrac{1}{4}E_C$ as was previously observed for ordered arrays of JJ \cite{Tighe1993,Delsing1994}. This further suggests that for $H\gg H^*$ the saturation is associated with charge fluctuations. Examination of the samples in the reverse direction of annealing (S2 to S1 to S0), we suggest that we observe a smooth crossover of macroscopic quantum effect from a regime dominated by Josephson coupling to a regime dominated by Coulomb interaction, which for each realization of coupling (annealing) and magnetic field could be attributed to a single junction behavior in the respective regime. 

Starting from a single resistively-shunted-JJ with $E_J \gg E_C$, for a small bias current a phase difference is established along the junction resulting in a zero-voltage supercurrent. However, at very low temperatures where activation is exponentially small, quantum fluctuations of the phase induce quantum tunneling of the phase variable  \cite{Caldeira1983} and thus a finite voltage \cite{Schwartz1985,Martinis1987}. With increasing charging energy, and in the limit $E_J\ll E_C$, a dual situation is expected, where the electric charge on the junction capacitance, which conjugates to the junction's phase, results in Coulomb blockade and thus an insulating state. For a small bias voltage, quantum fluctuations of the charge may give rise to a coherent current of cooper pairs and to dissipation due to single electron tunneling \cite{Schon1990}. The latter is also a macroscopic quantum effect due to the participation of the collective electron system in the process \cite{Iansiti1987,Geerligs1990}. 

To rationalize the analogy with a single resistively-shunted-JJ we follow Fisher \cite{Fisher1986} assuming that junctions in the $E_J\gg E_C$ regime (Samples S2 and S1) are resistively shunted with a wide distribution of shunt resistors. While in general the idea of a normal conductance channel was challenged in the limit of $T\ll T_c$ for an ordered array of JJs \cite{Chakravarty1987}, we believe that it is natural to expect normal regions in a random array of grains such as in our In/InOx composite, especially in the presence of magnetic field \cite{Hen2021,Zhang2021}. Using an Ambegaokar, Halperin and Langer approach \cite{Ambegaokar1971}, we assume that there exists a set of junctions which are coupled to shunt resistors large enough to allow phase slips and are connected to form an infinite network that spans the system. By the nature of the effect where the phase slips span the size of the sample for a magnetic field $H_{SM}\ll H_c^*$, we can map it on a random-resistor network (RRN) percolation problem, which at very low temperature, for a given tuning parameter (here the magnetic field) is self-organized to be at criticality. A well known result for percolation theory is that the resistance above percolation is bounded from below by the total resistance of the so-called ``singly connected bonds'' \cite{Coniglio1981,Coniglio1982,Wright1986}, that is, the bonds that if cut will disrupt the integrity of the infinite network. Near criticality, for a system of size $L$, the average conductance is given by $\langle G\rangle =L^{d-1}\langle g_1\rangle$, where $d$ is the dimensionality and $g_1$ is the linear conductance of the chain of singly connected bonds estimated here as $\langle g_1\rangle\sim 1/Lr_s$, where $r_s$ is the limiting shunt resistance \cite{EfrosShklovskii}. In two-dimensions, the value of the saturated resistivity within this RRN-percolation model is easily estimated as $\rho_s\approx \langle G\rangle^{-1} \approx r_s$. Thus, the regime of anomalous metal is a consequence of MQT through junctions characterized by $r_s\ll R_Q$.

Turning to the more insulating samples (S0, and high-field S1), the intergrain mediating InOx layer is insulating enough (single junction resistance $\gg R_Q$), to prevent an infinite cluster of percolating phase-coherent superconducting grains. In this regime, the small grains, which appear interstitially between the larger superconducting islands, become the bottleneck for continuous conduction due to their large charging energy satisfying $E_C\gg E_J$. This situation is amplified with increasing magnetic field where more of the weaker junctions between large superconducting islands lose phase coherence.  At the same time, a true insulating state as a consequence of Coulomb blockade may fail because of MQT of charge \cite{Geerligs1990,Delsing1994}. This dual behavior to the phase-MQT suggests to extend the percolation description according to Ambegaokar, Halperin and Langer results \cite{Ambegaokar1971} to the insulating side where a saturated resistivity is expected with limiting junction shunt resistance $r_s\gg R_Q$ and $E_C\gg E_J$. 

The phase-charge duality  is further understood by the intrinsic duality of the RRN of the randomly distributed shunt resistances underlying the JJ array. Indeed, there is an exact relation between the bulk effective conductivity of a 2D continuing composite made of two isotropic components and that of the dual composite \cite{Keller1964,Dykhne1970,Mendelson1975,Straley1977,Wright1986}. For a rectangular array of JJ representing bonds with resistances $\{r_i\}$ and conductances $\{g_i\}$ (where $g_i^D=1/g_i\equiv r_i$), the dual total conductance $G^D_x(g_1^D,g_2^D....g_N^D)$ along $x$-direction is related to the conductance along the $y$-direction $G_y(g_1,g_2, ...g_N)\equiv R_y(r_1,r_2,...r_N)$ as: $G_x^D=1/G_y$.  The same average behavior is expected for the random system, which is manifested in our random In/InOx composite \cite{Keller1964,Dykhne1970,Wright1986}: $\langle G_x^D\rangle =\langle R_y\rangle$. The charge-phase correspondence maps onto the RRN duality and further reinforces the vestiges of duality that originate from the H-SIT as previously suggested by Shimshoni {\it et al.}, \cite{Shimshoni1998}.

A further consequence of the above model is the realization of the fragility of the superconducting or insulating states from which the failed superconductor and failed insulator emerge. Besides the annealing and magnetic field, any perturbation applied to the system strongly affects the distribution of junctions properties, thus leading to a shift in the potential barrier for the MQT process resulting in a shift of the saturated resistance.  For example, external radiation can work to enhance the quantum fluctuations and thus AMS will emerge with larger saturated resistance as $T\to 0$ \cite{Tamir2019}, or reduce them thus pin the superconducting state \cite{Zhang2021}. Similarly, a ground plane near a sample can provide another dissipation channel to pin the superconducting state, as well as alter the charging energies to modify the saturation on the insulating side, both effects were observed in JJ arrays \cite{Rimberg1997} and in highly disordered films  \cite{Mason2002}.

In summary, we identify two regimes in anomalous metallic phase -- a ``failed superconductor'' and a ``failed insulator,'' as clearly suggested by resistivity saturation on both sides of an ``avoided'' SIT. The close connection between our granular composite and Josephson junction arrays is revealed by a set of empirical fit to the temperature dependence of resistivity. Quantum fluctuations of phase drive the transition from superconductor to anomalous metal when superconductors fail to establish global phase coherence, while quantum fluctuations of charge number drive the transition from insulator to anomalous metal when charge localization gives way to dissipative transport. A duality picture is strongly justified in our discussions by interchanging the roles of $N$ and $\phi$ and corroborated by the resemblance of transport behaviors when exchanging the roles of $\rho$ and $\sigma$. 


\noindent {\bf Acknowledgements:} We acknowledge discussions with Steven Kivelson, Sri Raghu, Boris Spivak, and Yacov Kantor. Work at Stanford University was supported by the National Science Foundation Grant NSF-DMR-1808385. Work at Tel-Aviv University was supported by the US-Israel Binational Science Foundation (Grant No. 2014098). We thank Sejoon Lim for assistance with SEM. Part of this work was performed at the Stanford Nano Shared Facilities (SNSF), supported by the National Science Foundation under award ECCS-1542152.

\bibliography{metal2}

\newpage

\onecolumngrid
\newpage
\setcounter{section}{0}
\setcounter{figure}{0}
\renewcommand{\thefigure}{S\arabic{figure}}
\renewcommand{\theequation}{S.\arabic{equation}}
\renewcommand{\thetable}{S\arabic{table}}
\renewcommand{\thesection}{S\arabic{section}}

\renewcommand{\thefootnote}{\fnsymbol{footnote}}

\begin{center}
\textbf{ SUPPLEMENTARY INFORMATION}

\vspace{3em}
\textbf{Anomalous metals: From ``failed superconductor'' to ``failed insulator''}\\

\fontsize{9}{12}\selectfont

\vspace{3em}
Xinyang Zhang,$^{1,2}$ Alexander Palevski,$^{3}$ and Aharon Kapitulnik,$^{1, 2, 4}$\\
\vspace{1em}
$^1${\it Geballe Laboratory for Advanced Materials, Stanford University, Stanford, CA 94305, USA}\\
$^2${\it Department of Applied Physics, Stanford University, Stanford, CA 94305, USA}\\
$^3${\it School of Physics and Astronomy, Raymond and Beverly Sackler, Faculty of Exact Sciences, Tel Aviv University, Tel Aviv 6997801, Israel}\\
$^4${\it Department of Physics, Stanford University, Stanford, CA 94305, USA.}\\
\end{center}

\vspace{20em}


\section*{Grain size distribution analysis}
\begin{figure}[ht]
	\centering
	\subfigure{\label{figs1:a}\includegraphics[width=0.3\columnwidth]{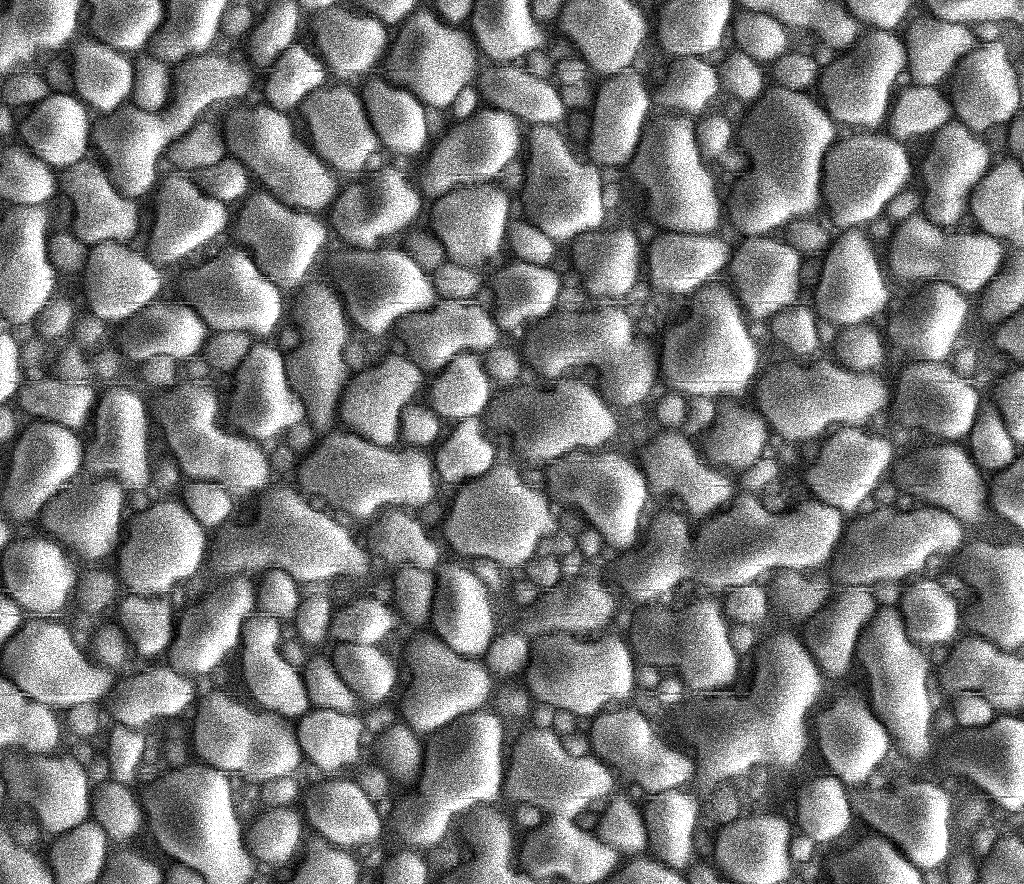}}
	\subfigure{\label{figs1:b}\includegraphics[width=0.3\columnwidth]{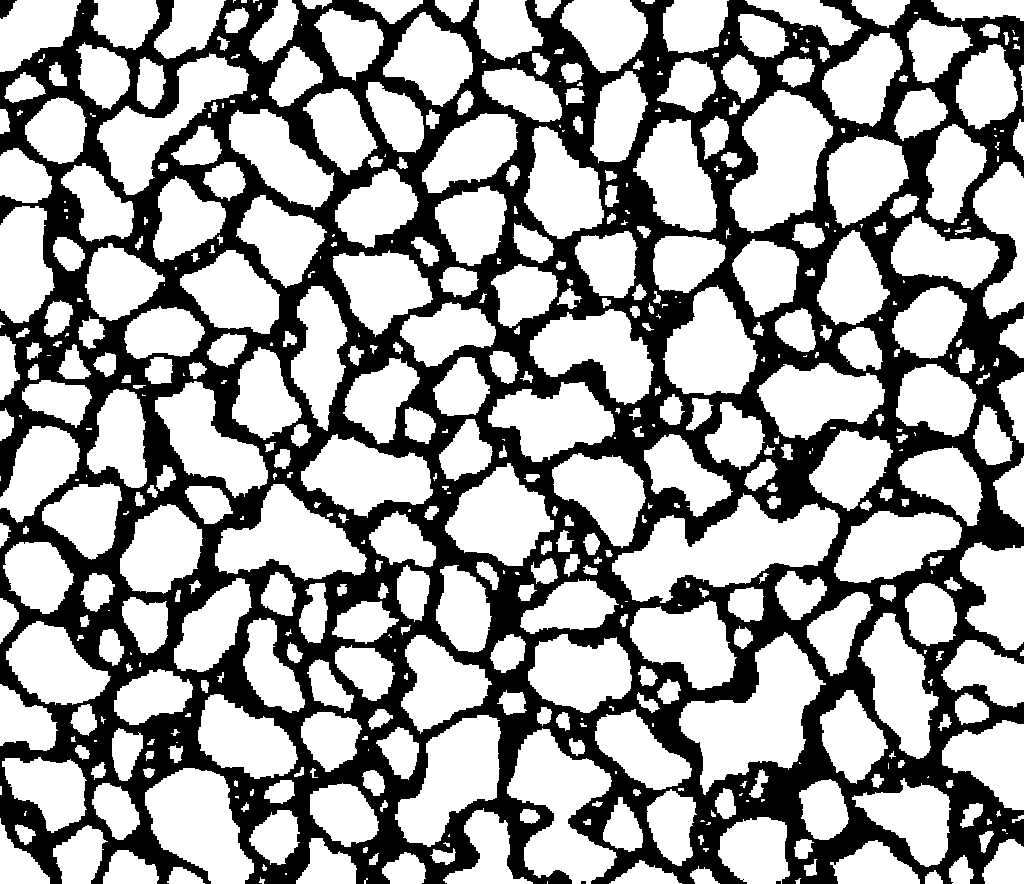}}
	\subfigure{\label{figs1:c}\includegraphics[width=0.3\columnwidth]{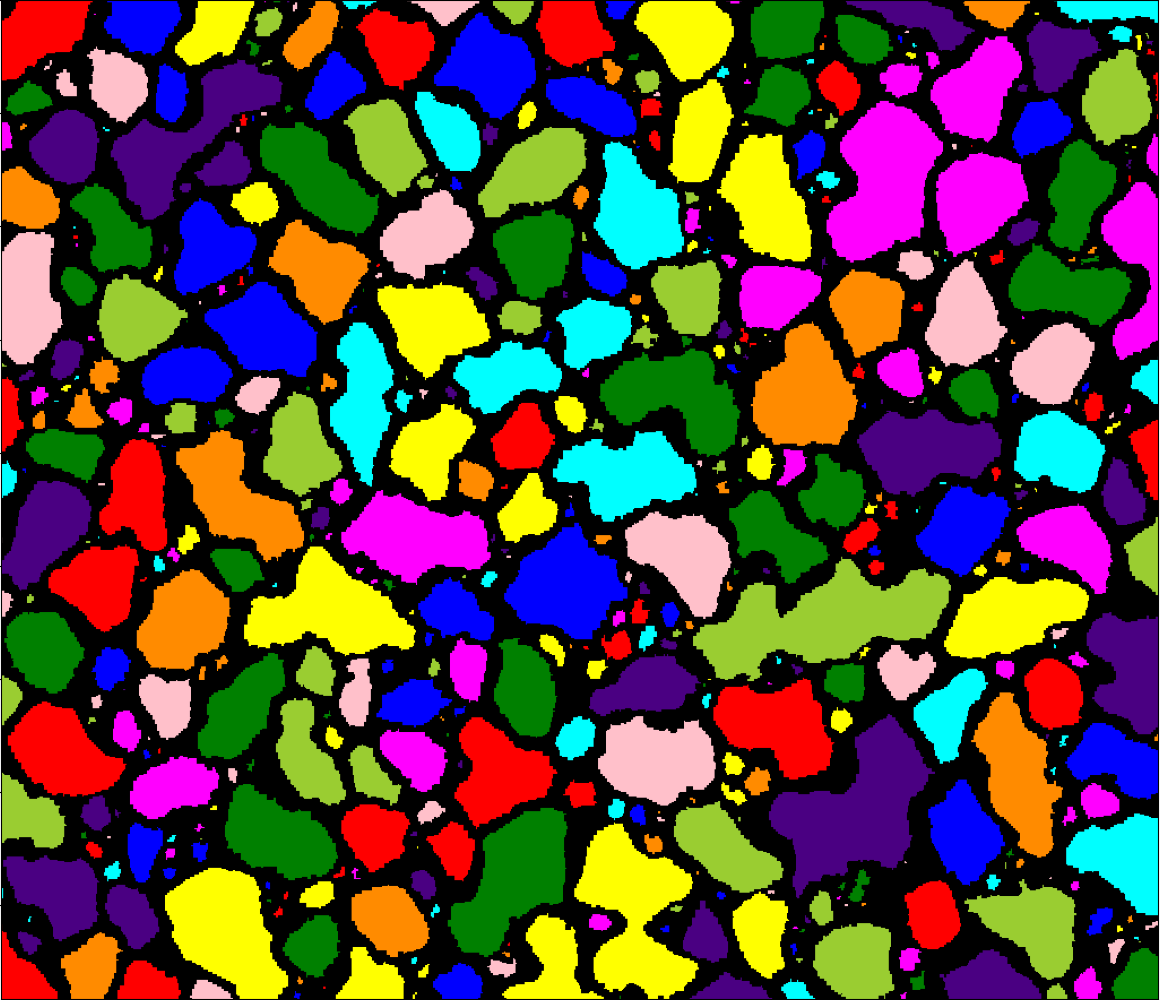}}
	\caption{\textbf{Grain clustering analysis on scanning electron microscopy (SEM) image.} (Left) The raw SEM micrograph. (Middle) Processed image in which white color represents grain occupation. (Right) Clustering outcome using scipy and opencv package in Python. The entire image is 6 $\mu$m wide. Each pixel corresponds to 6 nm.}
	\label{figs1}
\end{figure}
Figure~\ref{figs1} shows the same SEM micrograph as shown in Figure 2(a) in main text. The illuminated grains are randomly distributed indium (In) islands while a uniform amorphous indium oxide (InO$_x$) thin film underlies the grains. Because of the non-wetting characteristic of In on InO$_x$, the grains form shapes of droplets, including larger ones on the order of $\sim\mu$m and small interstitial ones that are $\sim$ nm in diameter. Therefore, the appearances of grains in this SEM picture is affected by the angle of illumination, and thus the dark side of the grains may be hard for image processing algorithm to identify.

Nevertheless, the edges of the grains have sufficient contrast to be delineated manually. While it is straightforward to distinguish the larger grains, the smaller interstitial ones are often blurred and not well contrasted from the background. Furthermore, since the diameter of the smallest grains are approaching image resolution, a grain that is smaller than a few pixel squared may not be picked up at all. As a consequence, we emphasize a lower cut-off in the distribution of measured grain equivalent diameter as shown in Figure 2(b) in main text. A similar approach was previously used to analyze percolation characteristics in inhomogeneous Pb deposited on amorphous Ge \cite{Kapitulnik1984}.

All isolated grains are identified using clustering algorithm in ``scipy'' package in Python. All pixels that belong to an isolated grain are collected with the same label and the location of each grain's centroid and the respective equivalent diameter can also be easily found. With this analysis, all geometric information in this picture regarding the random distribution of In grains is collected. Further analysis may reveal the distribution of intergrain distance, spacing, or coordination numbers of the grains, etc. These information may be useful in studying intergrain coupling or tunneling processes in the system.\\

\section*{Low-field magnetoresistance}
\begin{figure}[ht]
	\centering
	\subfigure{\label{figs2:a}\includegraphics[width=0.47\columnwidth]{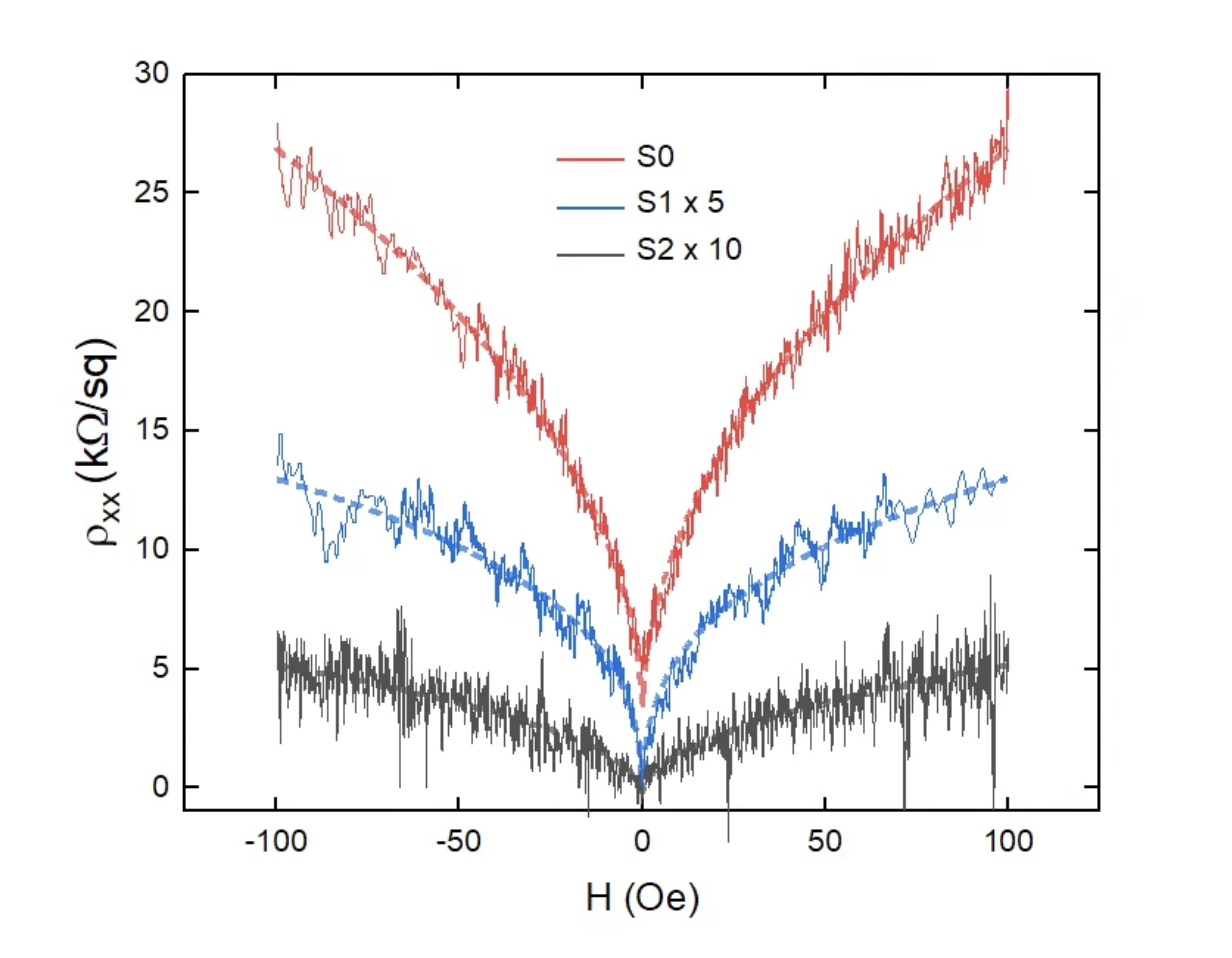}}
	\subfigure{\label{figs2:b}\includegraphics[width=0.45\columnwidth]{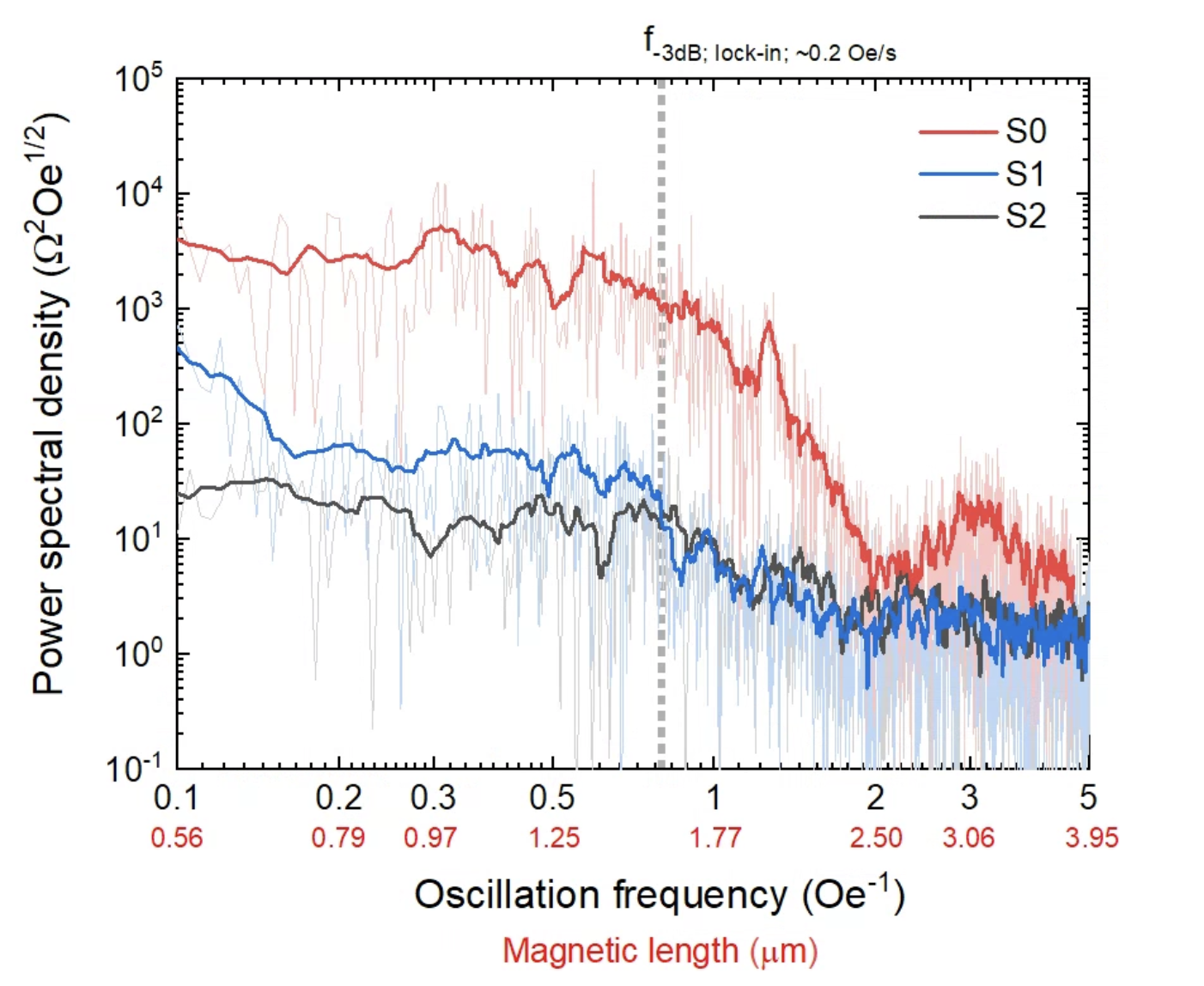}}
	\caption{\textbf{Magneto-resistance and its fluctuation in S0, S1, and S2.} (Left) Magnetoresistance (MR) between $H=\pm 100$ Oe. Note the difference in scale for S1 and S2. Dashed lines are power-law fits that remove general trend of the MR. (Right) Power spectral density as a function of Fourier component of magnetic field fluctuation. ``f$_{3\text{dB}}$'' denotes the cut-off frequency of our lock-in amplifier's low pass filter in this measurement.}
	\label{figs2}
\end{figure}

Figure~\ref{figs2} shows a comparison of MR between three annealing stages S0, S1, and S2. Dashed lines are fits to the MR using power law form $\rho_{\text{xx}}=\rho_0+A\cdot|H|^{p}$. In additional to an overall positive MR, the fluctuations may result from persistent current loops connected by inter-grain Josephson coupling. In certain cases, loops of a prominent size may appear in the Fourier spectrum with a peak at a certain frequency. However, in our system due to the randomness, current loops of all sizes may contribute and the resulting spectrum is that of a white noise, i.e. without apparent frequency dependence. Here, we see general spectra of white noise after taking the -24 dB/octave lock-in low-pass filter cut-off at $\sim$ 0.8 $Oe^{-1}$ into account. For S0, the peaks at around 1.3 $Oe^{-1}$ and 3 $Oe^{-1}$ may originate from current loops of $2$ or $3$ $\mu$m in size defined as $\ell_B=\sqrt{\Phi_0/2\pi H}$, corresponding to the large islands. $\Phi_0=2.07\ T\cdot m^2$ denotes a flux quantum. For S1 and S2, the spectra show no prominent peaks suggesting current loops of all sizes contribute to frequency-independent spectra. This is consistent with expectation as superconductivity is more robust in S1 and S2, where a low-field is not sufficient to cut off a majority of the Josephson couplings. The absolute values of the spectral density, however, may not matter since they depend on sensitivity settings of lock-in amplifier as S0, S1, and S2 have vastly different resistance values. Where periodicity is inherent in the morphology of the superconducting islands array, a periodic MR oscillations are clearly observed as in Ref.~\cite{Bottcher2018} and Ref.~\cite{Yang2019}.\\

\section*{Differential resistivity $dV/dI$}
\begin{figure}[ht]
	\centering
	\includegraphics[width=0.5\columnwidth]{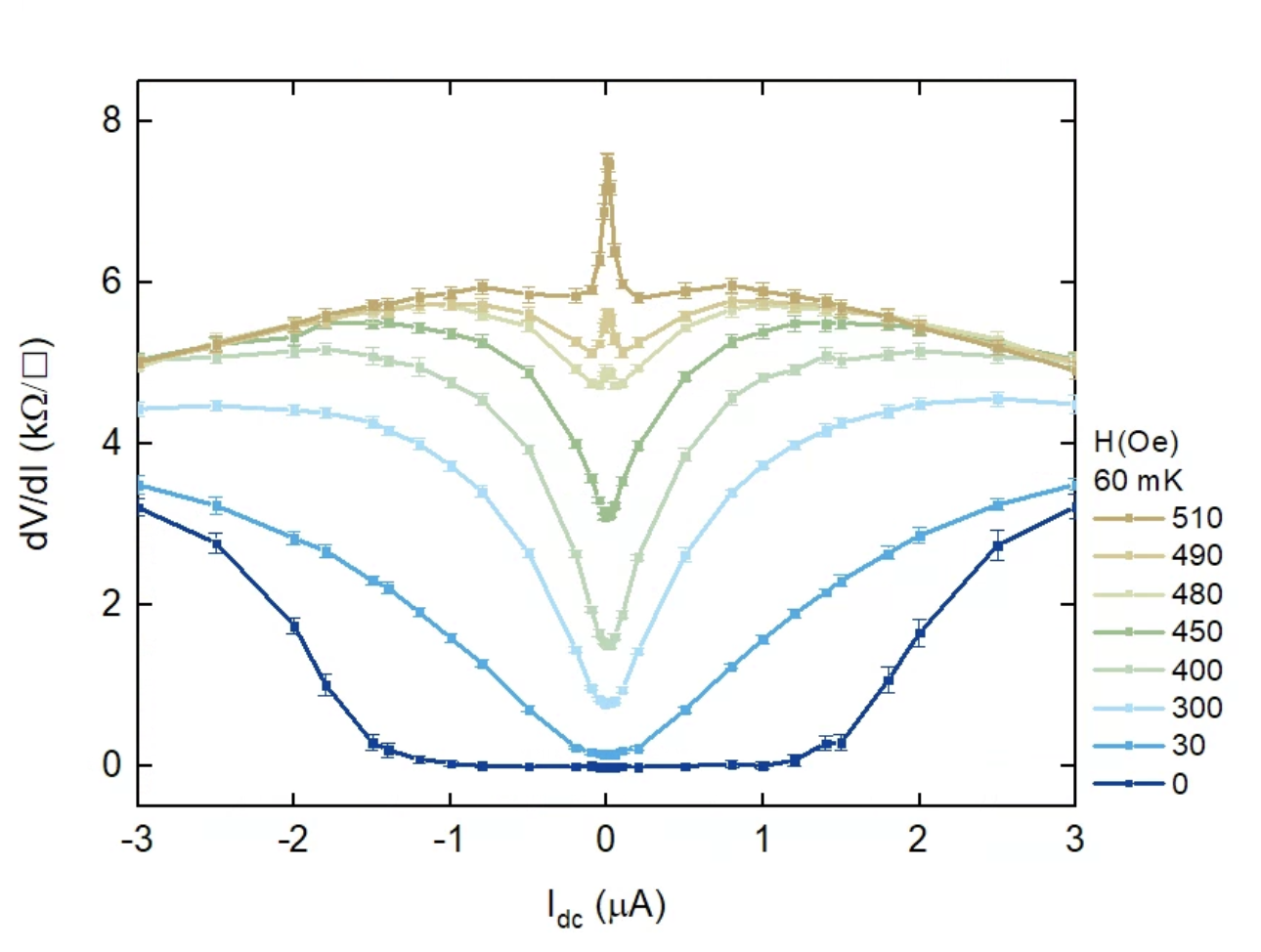} 
	\caption{\textbf{Differential resistivity $dV/dI$ in different regimes.} $dV/dI$ versus dc bias current $I_{dc}$ in superconducting, failed superconducting, and failed insulating regimes, as controlled by magnetic field.}
	\label{figs3}
\end{figure}

Figure~\ref{figs3} shows differential resistivity $dV/dI$ measured at fixed dc bias current for different magnetic field. In this case, an applied magnetic field tunes the system from superconducting to failed superconducting to failed insulating phases. The critical current in superconducting phase is immediately suppressed upon a low applied field of 30 Oe. In the failed superconducting phase, $dV/dI$ has a v-shape, with its value approaching the normal state value at higher currents. The v-shape gradually evolve to a zero-bias peak, where the system enters failed insulating phase. The peak grows rapidly as field increases and eventually diminishes when local superconductivity amplitude is suppressed at high field. Such I-V characteristics are also observed as a feature of anomalous metal in ordered array of JJ as in Ref.~\cite{Bottcher2018}.

\end{document}